\documentclass[manuscript,nonacm]{acmart}
\AtBeginDocument{%
  \providecommand\BibTeX{{%
    \normalfont B\kern-0.5em{\scshape i\kern-0.25em b}\kern-0.8em\TeX}}}

\begin{document}

\title{Shots and Boosters: Exploring the Use of Combined Prebunking Interventions to Raise Critical Thinking and Create Long-Term Protection Against Misinformation}
\begingroup
\renewcommand\thefootnote{}\footnote{
 This paper was presented at the 2025 ACM Workshop on Human-AI Interaction for Augmented Reasoning (AIREASONING-2025-01). This is the authors’ version for arXiv.}
\endgroup

\author{Huiyun Tang}
\affiliation{
  \institution{University of Luxembourg}
  \city{Esch-sur-Alzette}
  \country{Luxembourg}}
\email{huiyun.tang@uni.lu}

\author{Anastasia Sergeeva}
\affiliation{
  \institution{University of Luxembourg}
  \city{Esch-sur-Alzette}
  \country{Luxembourg}}
\email{anastasia.sergeeva@uni.lu}



\renewcommand{\shortauthors}{Tang et al.}

\begin{abstract}
The problem of how to effectively mitigate the flow of misinformation remains a significant challenge. The classical approach to this is public disapproval of claims or "debunking". The approach is still widely used on social media, but it has some severe limitations in terms of applicability and efficiency. An alternative strategy is to enhance individuals' critical thinking through educational interventions. Instead of merely disproving misinformation, these approaches aim to strengthen users’ reasoning skills, enabling them to evaluate and reject false information independently. In this position paper, we explore a combination of intervention methods designed to improve critical thinking in the context of online media consumption. We highlight the role of AI in supporting different stages of these interventions and present a design concept that integrates AI-driven strategies to foster critical reasoning and media literacy.
\end{abstract}




\maketitle

\section{Introduction}\label{sec:introduction}
The fast spread of misinformation is an enormous challenge for both society and individuals, with a great impact on democracy \cite{freeze2021fake}. In light of the grave consequences, various digital misinformation interventions have been implemented to slow its propagation. The most common way to combat misinformation is fact-checking. Traditionally, fact-checking was performed by journalists \cite{graves2019fact}, initially as part of the internal validation process (meaning that if the information in the paper was not correct, the paper would never be published), and later externally, for the broader audience through fact-checking websites \cite{graves2019fact} and collaborations with social networks \cite{micallef2022true}. Another form of fact-checking is the crowdsourcing model. Mostly known for its implementation on X as "Community Notes", the crowdsourcing model relies on ordinary users who provide their insights on the veracity of information \cite{chuai2024community}.
Finally, to achieve better scalability and coverage of social networks, there is a growing number of studies dedicated to the use of AI optimisation, both in the fact-checking/identifying questionable content \cite{guo2022survey} and in informing users about the results of this assessment\cite{horne2019rating,lai2019human,pareek2024effect}.
These interventions can be broadly assigned to the "debunking" type of interventions\cite{ecker2022psychological}. The goal of debunking interventions is to inform the user that the content they are reading contains incorrect information, sometimes explaining exactly what is wrong.

Some limitations have been reported on the effectiveness of the debunking approach. A primary concern is a limited scope: only a fraction of misinformation undergoes verification\cite{epstein2020will}.
Crowdsourcing and AI-driven approaches can alleviate these problems by expanding the scale of fact-checking; however, misinformation usually spreads very quickly, which creates a significant time gap between initial spread and debunking, affecting its effectiveness \cite{chuai2024did}. In addition, exposure to misinformation has lasting effects, making it challenging to mitigate it by debunking\cite{lewandowsky2012misinformation}.
Given these challenges, relying only on debunking is not enough. 
We also need to improve people's critical thinking skills and help them resist misinformation more effectively.

Currently, there are two main approaches aimed at raising people's critical thinking and improving their skills in identifying misinformation: nudging and inoculation. Nudging involves subtle design interventions designed to trigger analytical thinking, urging users to engage their critical thinking skills\cite{kozyreva2024toolbox}. For example, introducing friction, such as asking people to pause before sharing information, activates their existing ability to critically evaluate content\cite{fazio2020pausing}. However, nudging alone does not directly correct existing misconceptions or improve people's skills; instead, it focuses on prompting them to use their existing abilities. This approach works only if the person already has all the necessary knowledge to recognize misinformation but just needs to be put in the right state of mind to apply it.

On the other hand, inoculation is an educational approach that aims to enhance or build analytical skills and media literacy. 
This approach is based on inoculation theory \cite{compton2021inoculation}, which suggests that exposing people to weaker forms of misinformation can help them develop mental resistance to stronger misinformation in the future\cite{prebin2024}. The approach includes different educational interventions, such as 
educational articles, videos and games\cite{1roozenbeek2019fake,dame2022combating}. Although those methods have shown promise, their effectiveness often diminishes without regular reinforcement. Recent advancements in AI present new opportunities to address these challenges and improve the inoculation approach. AI-driven educational tools can deliver personalized learning experiences tailored to individual users’ needs, offer immediate and detailed feedback, and facilitate interactive experience that actively engage learners\cite{leong2024putting}. Several successful examples demonstrate the effectiveness of rapid interactions with AI agents in improving cognitive thinking, leading to immediate results\cite{tanprasert2024debate, danry2023don, liang2023encouraging}. At the same time, media literacy interventions and serious games (including games with AI-based elements) have been shown to be effective in improving long-term critical thinking skills, particularly in the context of media literacy \cite{tang2024mystery, tang2025breaking, roozenbeek2019fake}.



Researchers suggested that a single intervention may have only limited effects\cite{kozyreva2024toolbox}. In this position paper, we suggest that nudging and inoculation should be understood as complementary rather than interchangeable strategies. While nudging taps into immediate, short-term cognitive resources, inoculation contributes to deeper, longer-term growth in critical thinking and media literacy. To support this idea, we first reviewed current tools for fostering critical thinking, especially focusing on the context of media literacy. Then, we propose a design concept that aims to improve the literacy of an individual in the media.

\section{Related Work}\label{sec:Related Work}
\subsection{Critical Thinking and Digital and Media Literacy} 
Critical thinking lacks a universally accepted definition \cite{hatcher2021thinking}. Following HCI literature\cite{tanprasert2024debate}, this paper defines critical thinking as the ability to “evaluate arguments or propositions and make judgments that inform beliefs and guide actions(\cite{huitt1998critical}p. 1).” In the context of online media consumption, critical thinking skills are closely related to digital literacy and media literacy. Digital literacy, a narrower aspect of information literacy, emphasizes the competencies in using digital tools and discerning credible sources\cite{bawden2008origins}. Media literacy focuses specifically on equipping users to detect biases and underlying messages, defined as the ability to “access, analyse, and produce information for specific outcomes” (\cite{aufderheide2018media}, p. 6). 

\subsection{Developing and Supporting Critical Thinking in HCI}
HCI research has explored various methods to support and enhance critical thinking. For clarity, we categorize these methods into two groups based on their intended duration of impact: short-term effects and medium-to-long-term effects.
\subsubsection{Short-term Effects interventions}
AI-powered tools and LLM have been extensively explored to foster critical thinking, particularly through simulated debates, personalized feedback, and adaptive interactions. For example, Tanprasert et al. developed an LLM-based chatbot capable of adopting distinct debating personas: "persuasive" (rational, understanding) and "eristic" (aggressive, confrontational) to encourage users to critically evaluate content encountered within YouTube's filter bubbles. Their results indicated that the persuasive chatbot was particularly effective in provoking deeper reflection on arguments\cite{tanprasert2024debate}. Similarly, Danry et al. employed an AI-based questioning framework inspired by the Socratic method, using targeted questions to stimulate critical reasoning and logical evaluation. Their findings showed significant improvements in the analytical engagement and logical discernment of users\cite{danry2023don}. Further research has explored the use of multiple LLM-based agents to debate opposing points of view, thus fostering divergent reasoning and encouraging users to critically reflect on alternative points of view\cite{liang2023encouraging}.

Another related approach involves real-time assessment and explanatory feedback to encourage analytical thinking when users interact with news content. Zavolokina et al. proposed an AI-driven digital nudging tool designed to shift users from intuitive, rapid thinking (System 1) to more deliberate, analytical thinking (System 2). Their intervention provided real-time propaganda detection along with AI-generated explanations, effectively slowing impulsive news consumption and improving users' ability to critically assess media content\cite{zavolokina2024think}.

\subsubsection{Medium-to-long-term Effects Interventions}

The second group of interventions include the various educational interventions (mostly in the context of media literacy education), aimed at raising people's awareness of misinformation and deepening their understanding of it. We identified two areas where AI can be useful for improving the education process: personalized AI-based learning courses and serious games with AI components. 

Jian \cite{jian2023personalized} specifies four areas where AI can be
used in the educational process: data analysis of student interactions, AI-powered virtual assistants for virtual tutoring, AI-based adaptive assessment (tailored for the students' needs) and adaptive resource recommender systems. In the context of developing critical thinking skills, AI-based characters can take the role of co-learners or tutors to lead user progression through learning materials \cite{pataranutaporn2021ai}, fostering reflection and making the process of acquiring knowledge more enjoyable. Similarly, \cite{binhammad2024investigating} showed that despite some difficulties in generating tailored content via AI, it is a prospective method for future educational intervention development.
In the domain of Media Studies, Parra-Valencia and Massey \cite{parra2023leveraging} pointed to the benefits of using AI agents to enhance digital literacy (which is often considered in relation to media literacy) and to promote access to information and resources, which, in turn, could help create a knowledge base, which is a core component of critical thinking processes \cite{lai2011critical}.

One of the potential ways to incorporate the AI components into a tailored intervention can be done via a serious games approach. It has a long story in the misinformation education domain with a goal to improve information literacy (e.g. \cite{roozenbeek2019fake}) via interactive and immersive/role-playing experiences. For instance, in games like Bad News \cite{roozenbeek2019fake}, Harmony Square \cite{harmonysquare}, Cat Park \cite{Gusmanson.nl_2022}, and ChamberBreaker \cite{jeon2021chamberbreaker}, players assume the role of misinformation creators, learning firsthand about deceptive tactics. Conversely, games such as MathE \cite{katsaounidou2019mathe}and Escape the Fake \cite{paraschivoiu2021escape} position players as detective-debunker, identifying fake news using verification tools such as reverse image search. 
In terms of AI, there was an attempt to develop a role-playing game where players acted as fact-checkers, reviewing storylines generated by LLMS. The results showed that the game enabled players to critically engage in misinformation through investigative role-play \cite{tang2024mystery}. Similarly, Tang et al. created a misinformation game where two players compete to sway the AI-represented public opinion by either creating or debunking misinformation; the results showed the efficiency of this intervention in improving users' misinformation discrimination abilities and will to use critical thinking while interacting with media\cite{tang2025breaking}.

Summarizing the existing literature, we can conclude that, while there are several attempts to apply AI at the different levels of prebunking interventions, the attempts are still limited and do not combine the benefits of short- and long-term interventions.

\begin{figure}[htbp]
    \centering
    \includegraphics[width=10cm]{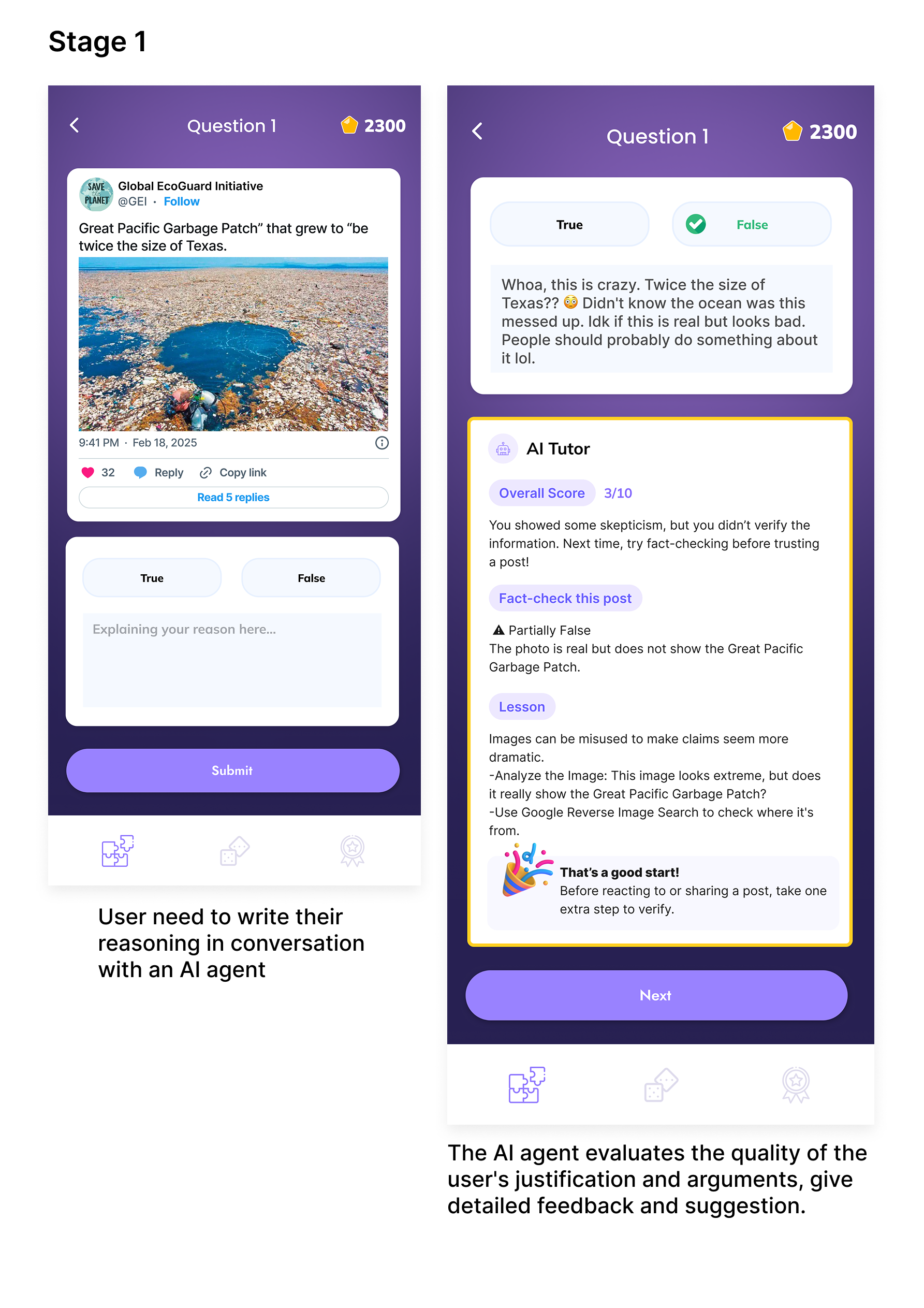}
    \caption{Presentation of Design concept}
    \label{fig:stage1}
\end{figure}

\begin{figure}[htbp]
    \centering
    \includegraphics[width=10cm]{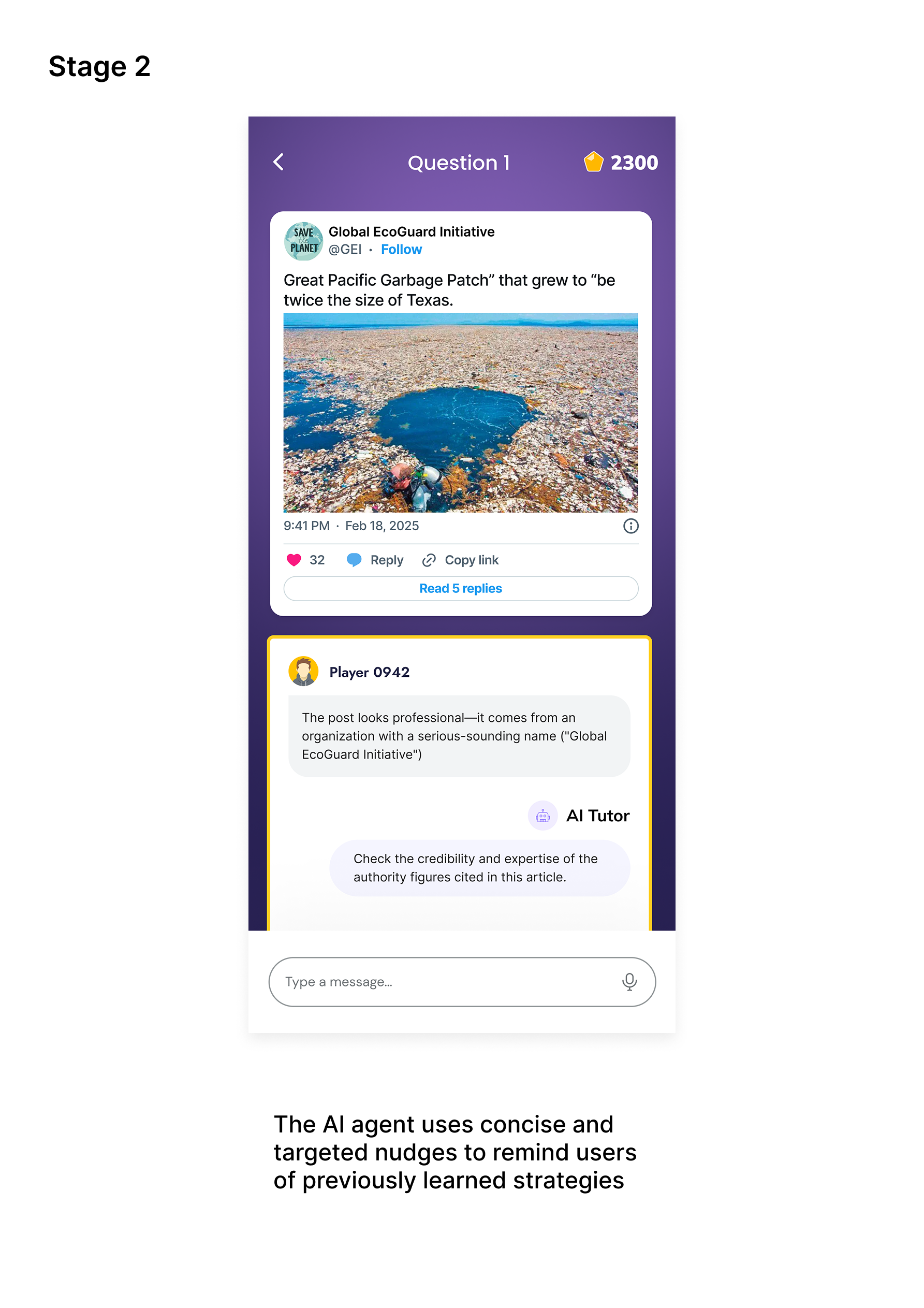}
    \caption{Presentation of Design concept}
    \label{fig:stage2}
\end{figure}

\section{Design Concept}\label{sec:Design concept}


Our proposed AI-driven intervention is designed as an interactive training game structured into two complementary stages, combining elements of inoculation theory and nudging. The training leverages AI's capacity for personalized interactions to enhance critical thinking and media literacy.

\subsection{\textbf{Stage 1: Inoculation Training ("Vaccination")}}

In the first stage, users participate in a simulated "vaccination" experience against misinformation through an interactive game. Players encounter various potentially misleading posts or news articles and must determine whether each is genuine information or not. The player also needs to write their reasoning in conversation with an AI agent, aiming to persuade the agent of their judgment. The AI agent acts as an intelligent tutor (it is also possible to implement multiple agents, who will take the roles of additional non-playable characters (NPC), helping users gather additional information). 
The AI agent evaluates the quality of the user's justification and arguments, guided by criteria drawn from professional journalistic practices \cite{EC2025CodeDisinformation}. Then, the agent provides an analysis of the user's arguments and corrects misconceptions. In case of low critical thinking and media literacy skills, the agent can provide a link to additional educational resources, which explains the basics of reasoning and information gathering; it can also show the example of a reasoning path, which can be applied to the gaming task (similarly to the rubric analysis of the Collegiate Learning Assessment Test \cite{CAE2014} ) 
Lastly, the agent can provide structural feedback for the user, including a summary or categorize the user's reasoning strategies, helping players become aware of their strengths and potential biases (see the proposed design concept in (\autoref{fig:stage1}). 

Through repeated interaction with the player, the AI agent learns the player's reasoning style, strengths, and weaknesses. This information will help the agent successfully teach the user, customize exercises, and adjust the difficulty level of the game. Additionally, the agent saves the user's strength and weakness profile for future "boosters" interventions. After achieving sufficient mastery (as indicated by consistent performance), the users move on to the second stage. 

\subsection{\textbf{Stage 2: Reinforcement Nudges ("Booster")}}

The second stage involves brief, personalized reinforcement Unlike previous stages, which involved extensive interactions with agents, this phase uses concise and targeted nudges delivered by the AI agent to remind users of previously learned strategies or correct any emerging misconceptions.

The main idea of this stage is to help users apply their new skills in everyday fact-checking situations. Therefore, the system should have the opportunity to incorporate examples from daily news or (based on the user's choice) the user's social media news feed and prompt the user to perform critical thinking exercises on this information in the form of short nudge-based exercises.

These nudges are dynamically generated and rooted in the AI agent's knowledge of each user's prior interactions. For example, if a user previously repeatedly made mistakes via trusting the source, which used authority claims (e.g. pretending to be medical authority) 
a follow-up nudge might remind them: "Check the credibility and expertise of the authority figures cited in this article." This brief and personalized reinforcement aims to retain the learned strategies over the longer term, maintaining cognitive immunity against misinformation with minimal cognitive load (See proposed design concept in (\autoref{fig:stage2})). 

\begin{acks}
The research was supported by the Luxembourg National Research Fund (REMEDIS,
REgulatory and other solutions to MitigatE online DISinformation
(INTER/FNRS/21/16554939)).
\end{acks}

\bibliographystyle{ACM-Reference-Format} 
\bibliography{Reference}

\end{document}